\documentclass[useAMS,usenatbib]{mnras}

\usepackage{hyperref}

\usepackage[usenames, dvipsnames]{color}

\usepackage{graphicx,amssymb,longtable,lscape} %rotating,
\usepackage[fleqn]{amsmath}
\usepackage{textcomp}
\usepackage{color}
\usepackage{multirow}
\usepackage{float}
\usepackage{caption}
\usepackage{times}
\usepackage{subcaption}
\usepackage{color}
\usepackage{soul}

\usepackage[english]{babel}
\usepackage{array}
\newcolumntype{L}[1]{>{\raggedright\let\newline\\\arraybackslash\hspace{0pt}}m{#1}}
\newcolumntype{C}[1]{>{\centering\let\newline\\\arraybackslash\hspace{0pt}}m{#1}}
\newcolumntype{R}[1]{>{\raggedleft\let\newline\\\arraybackslash\hspace{0pt}}m{#1}}

\def\apj{\mbox{ApJ}}
\def\apjl{\mbox{ApJL}}
\def\apjs{\mbox{ApJS}}

\def\mnras{\mbox{MNRAS}}
\def\aj{\mbox{AJ}}
\def\araa{\mbox{ARA\&A}}

\def\aap{\mbox{A\&A}}

\title[Mid-IR and CO gas properties of FeLoBAL quasar]
{The mid-infrared and CO gas properties of an extreme star-forming FeLoBAL quasar}

\author[Pitchford et al.]{Lura K. Pitchford$^1$, Duncan Farrah$^{2,3}$, Katherine Alatalo$^{4,5}$, Jos\'{e} Afonso$^{6,7}$,
\newauthor  Andreas Efstathiou$^8$, Evanthia Hatziminaoglou$^9$,
Mark Lacy$^{10}$, Tanya Urrutia$^{11}$,
\newauthor  Giulio Violino$^{12}$ \\
$^1$Department of Physics, Virginia Tech, Blacksburg, VA 24061, USA\\
$^2$Department of Physics and Astronomy, University of Hawaii, 2505 Correa Road, Honolulu, HI 96822, USA \\
$^3$Institute for Astronomy, 2680 Woodlawn Drive, University of Hawaii, Honolulu, HI 96822, USA \\
$^4$Space Telescope Science Institute, 3700 San Martin Drive, Baltimore, MD 21218, USA\\
$^5$The Observatories of the Carnegie Institution for Science, 813 Santa Barbara St., Pasadena, CA 91101\\
$^6$Instituto de Astrof\'{i}sica e Ci\^{e}ncias do Espa\c{c}o, Universidade de Lisboa, OAL, Tapada da Ajuda, PT1349-018 Lisboa, Portugal \\
$^7$Departamento de F\'{i}sica, Faculdade de Ci\^{e}ncias, Universidade de Lisboa, Edif\'{i}cio C8, Campo Grande, PT1740-016 Lisbon, Portugal \\
$^8$School of Sciences, European University Cyprus, Diogenes Street, Engomi, 1516 Nicosia, Cyprus \\
$^9$ESO, Karl-Schwarzschild-Str. 2, 85748 Garching bei M\"unchen, Germany\\ 
$^{10}$National Radio Astronomy Observatory, 520 Edgemont Road, Charlottesville, VA 22903, USA \\
$^{11}$Leibniz-Institut f\"{u}r Astrophysik Potsdam (AIP), An der Sternwarte 16, D-14482 ¨ Potsdam, Germany\\
$^{12}$Centre for Astrophysics Research, University of Hertfordshire, College Lane, Hatfield AL10 9AB, UK \\}

\begin{document}

\pagerange{\pageref{firstpage}--\pageref{lastpage}} \pubyear{2019}

\maketitle

\label{firstpage}

\begin{abstract}

We present a detailed study of a high-redshift iron low-ionization broad absorption line (FeLoBAL) quasar (SDSS1214 at $z = 1.046$), including new interferometric $^{12}$CO \mbox{\textit{J}=2--1} observations, optical through far-infrared photometry, and mid-infrared spectroscopy. The CO line is well-fit by a single Gaussian centered \mbox{40\,km\,s$^{-1}$} away from the systemic velocity and implies a total molecular gas mass of \mbox{$M_\textnormal{gas} = 7.3 \times 10^{10}\,\textnormal{M}_\odot$}. The infrared SED requires three components: an active galactic nucleus (AGN) torus, an AGN polar dust component, and a starburst. The starburst dominates the infrared emission with a luminosity of \mbox{log($L_\textnormal{SB}[\textnormal{L}_\odot]) = 12.91^{+0.02}_{-0.02}$}, implying a star formation rate of about \mbox{2000\,M$_{\odot}$yr$^{-1}$}, the highest known among FeLoBAL quasars. The AGN torus and polar dust components are less luminous, at \mbox{log($L_\textnormal{AGN}[\textnormal{L}_\odot]) = 12.36^{+0.14}_{-0.15}$} and \mbox{log($L_\textnormal{dust}[\textnormal{L}_\odot]) = 11.75^{+0.26}_{-0.46}$}, respectively. If all of the molecular gas is used to fuel the ongoing star formation, then the lower limit on the subsequent duration of the starburst is 40\,Myr. We do not find conclusive evidence that the AGN is affecting the CO gas reservoir. The properties of SDSS1214 are consistent with it representing the endpoint of an obscured starburst transitioning through a LoBAL phase to that of a classical quasar.

\end{abstract}


\section{Introduction}\label{intro}

Nuclear activity and star formation coexist in many infrared-luminous systems across all redshifts \citep[e.g.][]{farrah03, alexander05, shi09, hatziminaoglou10, mainieri11, harris16}. Further, the evolution from a gas-rich major merger to a gas-poor elliptical galaxy is often attributed to an intense star formation episode, which either coincides with or precedes a luminous active galactic nucleus (AGN) phase \citep[e.g.][]{sanders88, priddey03, omont03, granato04, hopkins08}. Most quasars are thus found in massive galaxies, which frequently exhibit ongoing or recent star formation \citep[as in e.g.][]{trump13}. CO observations of quasars also find large gas reservoirs of \mbox{$10^{8} - 10^{11}\,\textnormal{M}_\odot$}, capable of fuelling further star formation \citep[e.g.][]{coppin08, iono09, riechers11, brusa15}. 

Within star-forming systems, the star formation rate (SFR) and stellar mass follow a tight correlation, which is often referred to as the star formation `main sequence.' The star formation trigger mechanism may depend upon where galaxies lie in relation to the main sequence; it may also depend upon evolutionary stage \citep[see e.g.][]{ricci17}. Star formation for galaxies along the main sequence is likely caused by internal secular processes. However, within those galaxies that fall above the main sequence, the so-called starburst galaxies, star formation likely results from major mergers \citep{daddi10}. For AGN, \cite{treister12} have found that mergers are responsible for the most luminous AGN, while the less luminous AGN result from secular processes (as in e.g. \citealt{schawinski11, mullaney12}, but see also \citealt{villforth17, hatziminaoglou18}). Thus, it follows that the most luminous AGN might also be housed by the most intense starbursts; that is, the scaling relation between AGN activity and star formation might be dependent upon AGN luminosity. Some authors do find SFR and AGN luminosity to scale with one another \citep[e.g.][]{bonfield11, imanishi11, young14, delvecchio15}, but others find no such scaling \citep[e.g.][]{shao10, harrison14, ma15, pitchford16}. There is also evidence which supports a correlation between the two, but only over certain AGN luminosity, SFR, and redshift ranges \citep{harris16}. Thus, the nature of the relation between an AGN and star formation within its host remains unclear. 

An interesting subclass of quasar in which to study this relation is that of the broad absorption line (BAL) quasars \citep{lynds67, turnshek84}, since some types of BAL quasars are thought to be young quasars whose hosts harbor ongoing starbursts. BAL quasars exhibit broad, blueshifted absorption features in their ultraviolet spectra with outflow velocities of $\gtrsim0.1c$ \citep{weymann91, hall02, trump06}. There are three types of BAL quasar, separated by the absorption features seen in their ultraviolet (UV) spectra. High-ionization BAL (HiBAL) quasars show absorption in Ly$\alpha \hspace{0.025cm} \lambda$1216, N\begin{scriptsize}{V}\end{scriptsize}$\hspace{0.025cm} \lambda$1240, Si\begin{scriptsize}{IV}\end{scriptsize}$\hspace{0.025cm} \lambda$1394 and C\begin{scriptsize}{IV}\end{scriptsize}$\hspace{0.025cm} \lambda$1549. HiBALs represent about 85 per cent of the total BAL population. Low-ionization BAL (LoBAL) quasars display all of the absorption features seen in HiBALs, as well as absorption from low-ionization ions such as Al\begin{scriptsize}{III}\end{scriptsize}$\hspace{0.025cm} \lambda$1857 and Mg\begin{scriptsize}{II}\end{scriptsize}$\hspace{0.025cm} \lambda$2799. These objects represent around 15 per cent of the BAL population. The third class, iron LoBALs (FeLoBALs), are the rarest of the three and have all of properties of LoBALs plus absorption from excited levels of iron \citep{hazard87}.

There are, broadly speaking, two different models for the origins of these BAL outflows \citep[as in e.g.][]{allen11}. In the first, the orientation effect, an outflow is only visible for a fraction of a quasar's lifetime and/or over certain viewing angles \citep[e.g.][]{elvis00}. In the other, a BAL outflow is likely indicative of a young object; that is, a quasar with BAL features in its spectrum would be in an early stage of its lifetime in which it is still partly obscured by gas and dust. This second scenario is often applied to LoBAL and FeLoBAL quasars. Furthermore, FeLoBAL quasars have high infrared luminosities of \mbox{$10^{12}-10^{13.3}$\,L$_{\odot}$} \citep{farrah07a}, with a significant fraction of them harboring SFRs in excess of \mbox{100\,M$_{\odot}$yr$^{-1}$} \citep[e.g.][]{urrutia09, farrah10}. There is also some evidence that the BAL winds in FeLoBAL quasars are actively quenching star formation \citep{farrah12}, though \cite{violino16} do not find evidence supporting this idea\footnote{\cite{schulze17} study a sample of LoBAL quasars at similar luminosities and find no differences between the Eddington ratios of the LoBALs and their non-BAL counterparts, which also argues, indirectly, against such an evolutionary scenario.}. FeLoBAL quasars are therefore valuable laboratories for studying the properties of AGN and star formation occurring simultaneously within a system and may be sites to examine the effects of an AGN wind on host star formation. The CO properties of FeLoBAL quasars, however, remain almost completely unexplored, especially in systems with constraints on the host SFR. To our knowledge, there exists only one other CO observation of a high-redshift FeLoBAL quasar \citep{ohta07}.

In this paper, we present the first interferometric CO study of a high-redshift FeLoBAL quasar with mid-infrared spectroscopic constraints on its star formation rate. Our target, SDSS J121441.42-000137.8 (hereafter SDSS1214), at \mbox{$z = 1.046$} \citep{hewett10}, shows prominent BAL troughs in its rest-frame UV spectrum, indicative of a powerful AGN-driven wind \citep{hall02}. Moreover, SDSS1214 is currently the most luminous star-forming FeLoBAL known. Its mid-infrared (MIR) spectrum shows prominent polycyclic aromatic hydrocarbons (PAHs), implying a high ongoing SFR \citep{peeters04}. As with the majority of far-infrared (FIR) bright quasars, the FIR emission is likely also dominated by star formation \citep[e.g.][]{hatziminaoglou10, harris16}. We combine our new \mbox{$^{12}$CO \textit{J}=2--1} data taken by the Combined Array for Research in Millimeter-Wave Astronomy (CARMA) with an updated and comprehensive modelling of the mid- through far-infrared emission in a multi-wavelength case study to investigate the nature of star formation in an FeLoBAL host galaxy.

This paper is structured as follows. Sec. \ref{sec:observations} describes the various data, as well as the method for fitting the spectral energy distribution (SED) using more data and models than have been previously explored with SDSS1214. Sec. \ref{sec:results} provides some of the properties of the CO gas and the SED fit, which are further explored in Sec. \ref{sec:discussion}. Lastly, Sec. \ref{sec:conclusions} presents our conclusions. Throughout this work, we assume \mbox{$H_0 = 70$\,km\,s$^{-1}$\,Mpc$^{-1}$}, \mbox{$\Omega = 1$}, and \mbox{$\Omega_{\Lambda} = 0.7$}.

\section{Analysis} \label{sec:observations}

\subsection{CO observations}

\begin{figure*}
	\begin{center}
		\includegraphics[width=0.95\textwidth]{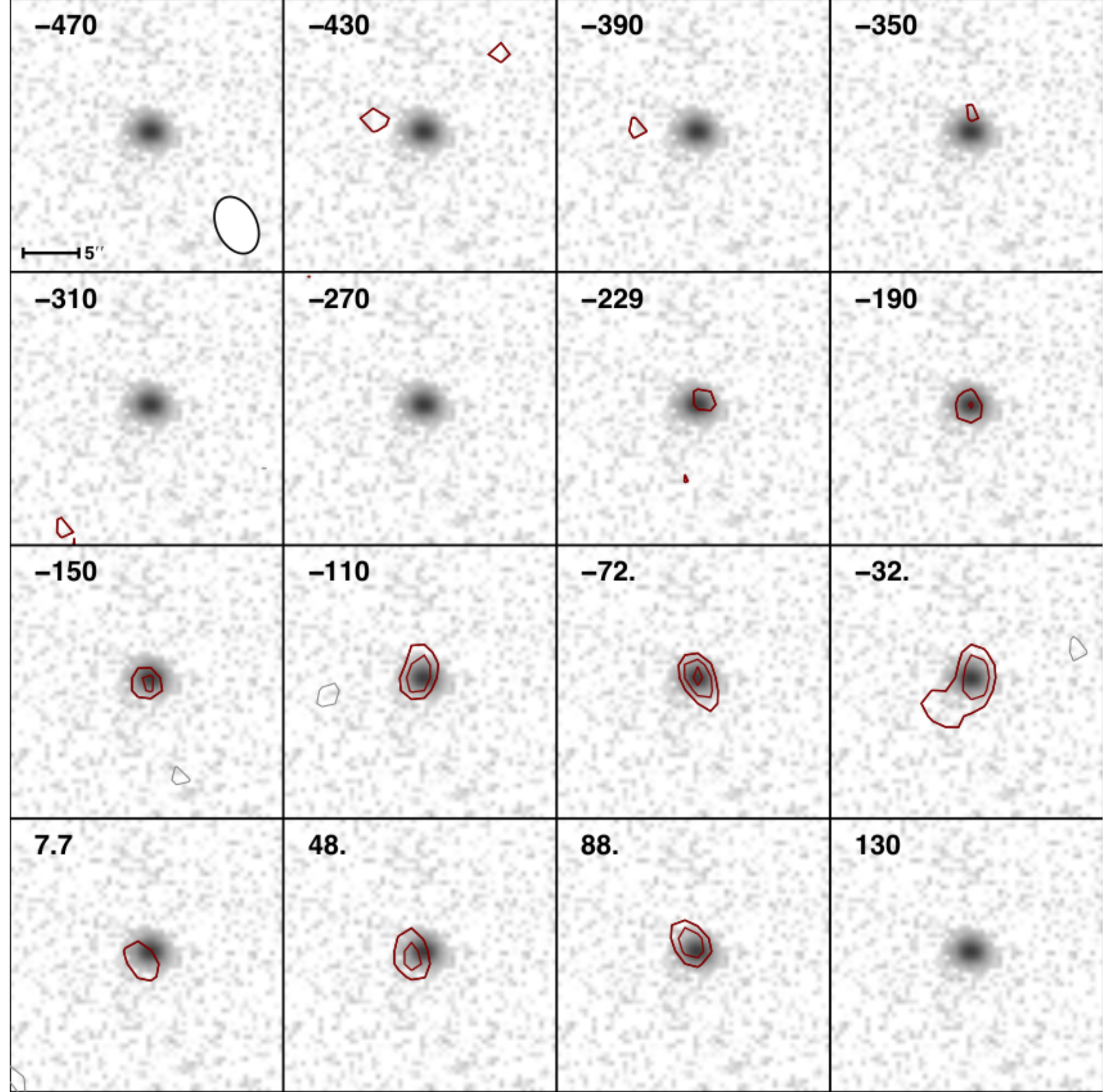}
		\caption{The CO channel maps overlaid on images constructed by summing the SDSS \textit{g}, \textit{r}, and \textit{i} bands. The number in the upper-left corner defines the central velocity of each channel, relative to the optically-defined systemic velocity of the quasar, and each channel is about \mbox{40\,km\,s$^{-1}$} in width. In the first panel, the ellipse in the lower-right corner shows the CARMA synthesized beam, while the bar represents 5", or 40\,kpc, at SDSS1214's redshift. The rms noise level is \mbox{1.4\,mJy\,beam$^{-1}$}. North is up, and east is to the left. The contours shown start at $\pm3\sigma$ and scale in $1\sigma$ increments. The negative contours are shown in grey. There seems to be more material at negative velocities in the immediate vicinity of the quasar when considering velocities of \mbox{-229--+88\,km\,s$^{-1}$}. However, these speeds are small enough to be attributable to the dynamics of the host galaxy, rather than indicative of a line-of-sight outflow.}
\label{fig:co_chanmaps}
\end{center}
\end{figure*}

The \mbox{$^{12}$CO \textit{J}=2--1} line from SDSS1214, with a rest frequency of \mbox{230.538\,GHz} (corresponding to a wavelength o \mbox{1.300\,mm}), was observed by CARMA\footnote{\href{http://www.mmarray.org}{http://www.mmarray.org}} between 18--26 May 2014, for a total on-source time of 19 hours. The CO line was observed in the D-array with baselines of \mbox{11-150\,m}, corresponding to a minimum spatial scale of $\sim$1.8" (\mbox{$\sim$15\,kpc}) at the frequency of the line. The first panel of Fig. \ref{fig:co_chanmaps} shows the CARMA beam with a FWHM of the synthesized beam of \mbox{5.4"$\times$ 3.6"}. The channel maps themselves will be discussed in Sec. \ref{sec:co_prop}.

SDSS1214 is located 4.3\,deg away from 3C273, which was used as a gain and passband calibrator. Data reduction was performed using the Multichannel Image Reconstruction, Image Analysis and Display (\textsc{miriad}) software package \citep{sault95} using standard methods identical to those described in Sec. 3.2 of \citet{alatalo13}.

\subsection{Optical spectrum}

\begin{figure}
	\begin{center}
		\includegraphics[width=7cm]{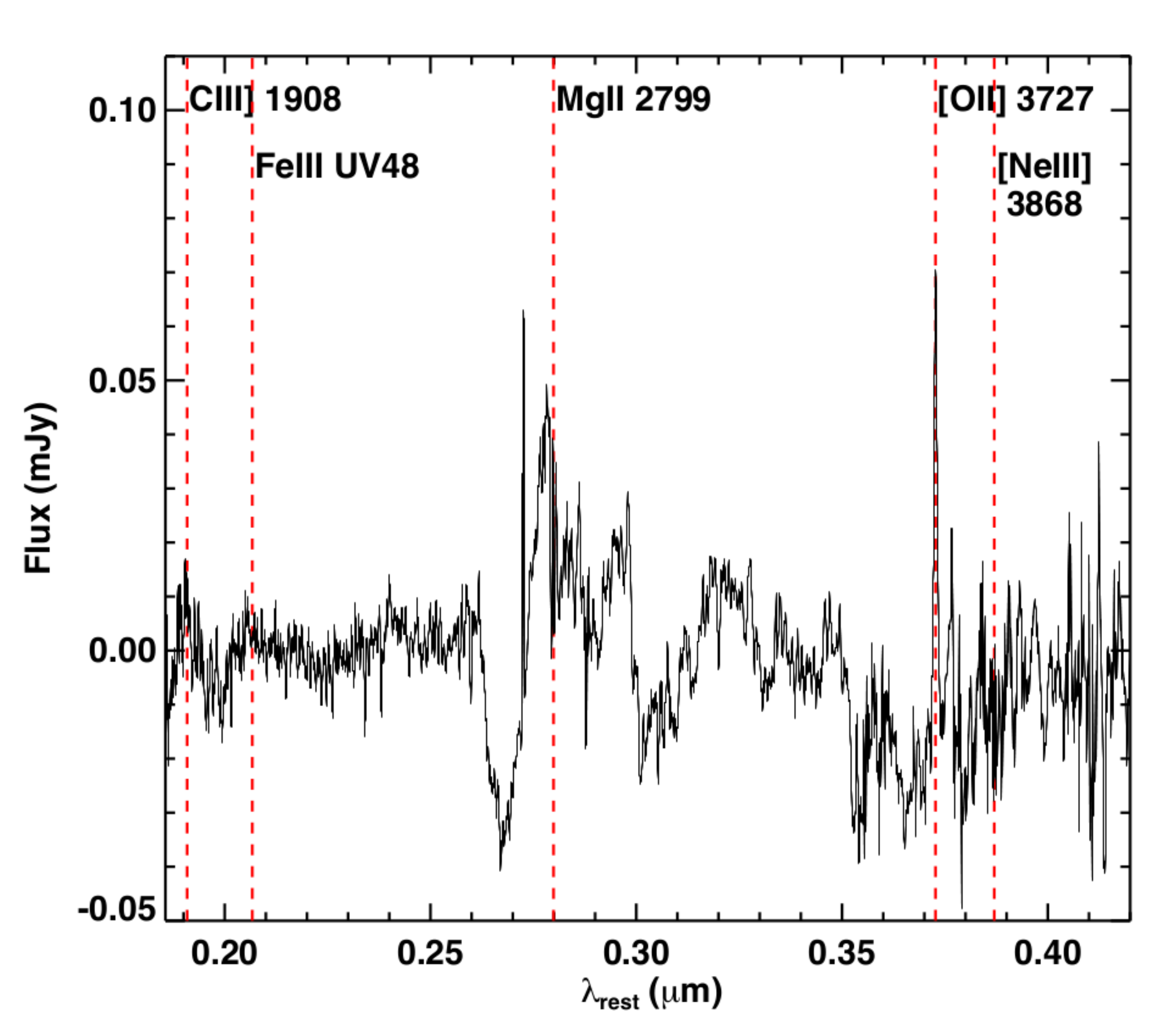}
		\caption{The optical data from SDSS. The lab-measured wavelengths of various emission lines, as well as FeIII UV48, which gives SDSS1214 its FeLoBAL distinction \protect\citep{hall02}, are shown as vertical, dashed lines. DR7 has a spectral resolution of \mbox{R $\sim 1800$} and a spatial resolution corresponding to 11\,kpc.}
\label{fig:optspec_spec}
\end{center}
\end{figure}

We present the optical spectrum in Fig. \ref{fig:optspec_spec}. These data come from the Sloan Digital Sky Survey \citep[SDSS, ][]{york00} Data Release 7 \citep[DR7;][]{abazajian09} with a spectral resolution of \mbox{R $\sim 1800$}. With a median PSF FWHM in the \textit{r}-band of 1.4" ($\sim 11$\,kpc at SDSS1214's redshift), the data are dominated by nuclear emission. 

The narrow [OII]$\lambda$3727 line present in the spectrum is used to define the redshift of \cite{hewett10}, \mbox{$z = 1.046$}. Throughout this work, whenever we reference the systemic velocity, we are referring to this narrow-optical defined redshift, as narrow, optical emission lines are less affected by the presence of an AGN than are other redshift indicators and likely trace the motion of the bulk of the ISM, thereby providing a redshift that is close to the true systemic value. 

Further, the optical spectrum shows a broad absorption feature in Mg\begin{scriptsize}{II}\end{scriptsize}$\hspace{0.025cm} \lambda$2799 with a weighted average velocity of \mbox{-13086\,km\,s$^{-1}$} \citep{trump06}. The presence of this feature indicates outflowing hot, ionized gas within the system and gives SDSS1214 its BAL classification.

\subsection{Infrared SED analysis}\label{sec:sed_analysis}

\begin{figure*}
	\begin{center}
        \includegraphics[width=3cm, height=3cm]{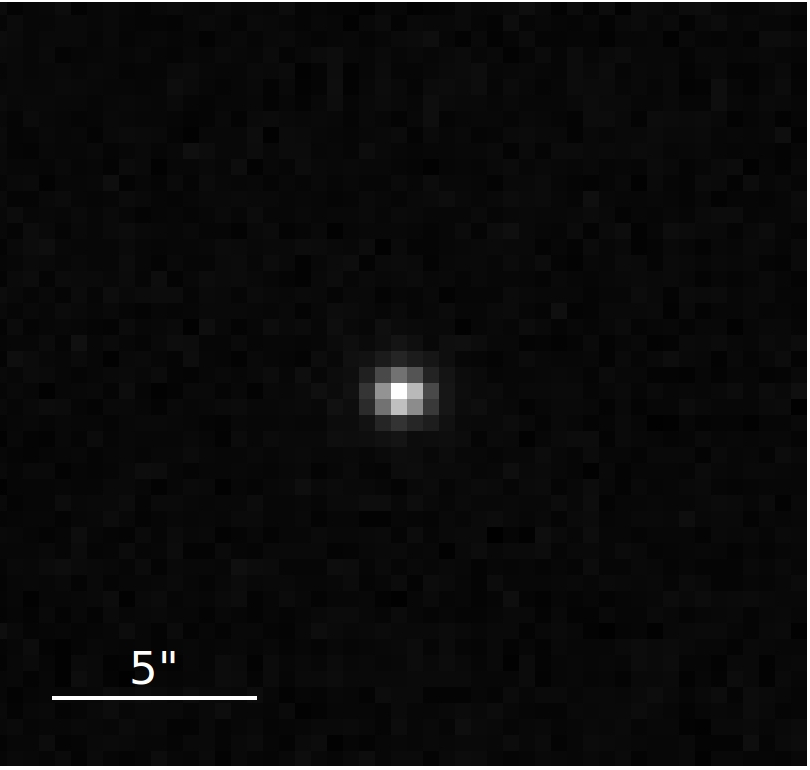} 
        \includegraphics[width=3cm, height=3cm]{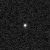}  
        \includegraphics[width=3cm, height=3cm]{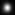}
        \includegraphics[width=3cm, height=3cm]{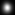}\\
        \includegraphics[width=3cm, height=3cm]{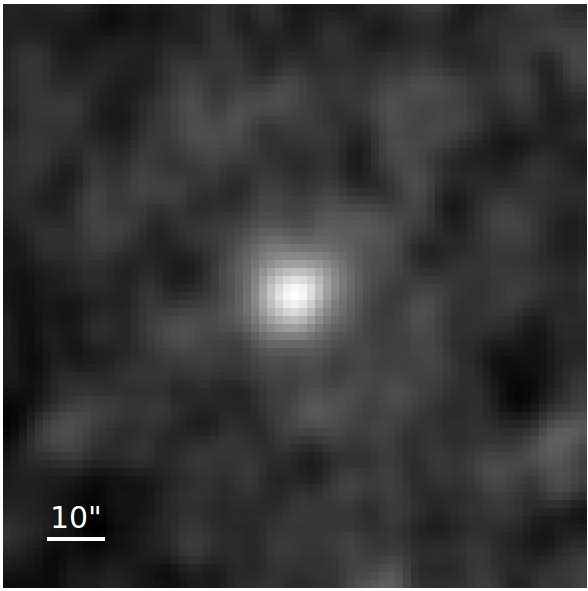}
        \includegraphics[width=3cm, height=3cm]{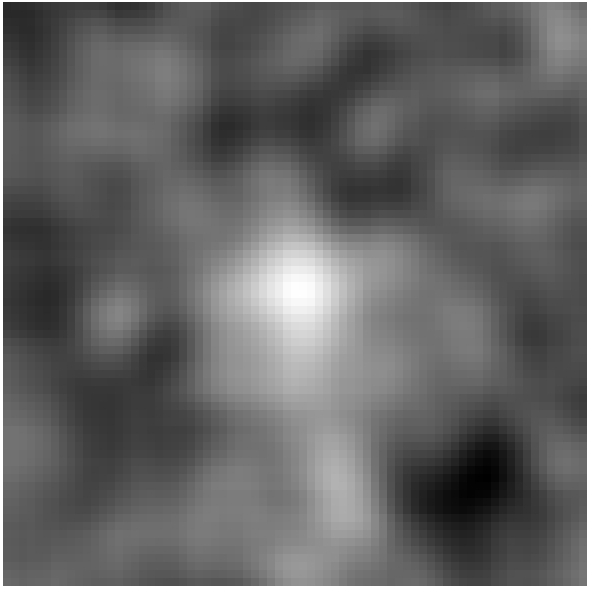} 
        \includegraphics[width=3cm, height=3cm]{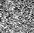}
        \includegraphics[width=3cm, height=3cm]{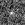}\\
        \includegraphics[width=3cm, height=3cm]{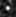} 
        \includegraphics[width=3cm, height=3cm]{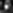}
        \includegraphics[width=3cm, height=3cm]{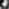} \\
        \caption{Cutouts from the SDSS \textit{i}-band, the UKIDSS \textit{K}-band, the four WISE bands, PACS 100 and 160$\mu$m, and the three SPIRE bands. North is up, and east is left. In the first row, each of the cutouts is 20" across, with the scale shown in the first panel. In the second and third rows, each cutout is 100" across with the first panel in the second row showing the scale. At SDSS1214's redshift, 1" corresponds to 8\,kpc.}
        \label{fig:cutouts}
    \end{center}
\end{figure*}

In addition to the optical photometry from SDSS, we have compiled all available infrared photometry. The near-infrared data come from the UKIRT Infrared Deep Sky Survey \citep[UKIDSS;][]{lawrence07}, while those in the MIR are from the Wide-field Infrared Survey Explorer \citep[WISE;][]{wright10} and the Multiband Imaging Photometer for \textit{Spitzer} \citep[MIPS;][]{rieke04}. Finally, the FIR data come from the \textit{Herschel} Astrophysical Terahertz Large Area Survey \citep[H-ATLAS;][]{valiante16}, which uses the Photodetector Array Camera and Spectrometer \citep[PACS;][]{poglitsch10} and the Spectral and Photometric Imaging Receiver \citep[SPIRE;][]{griffin10} to image across five bands. The two PACS bands, $100\,\mu$m and 160\,$\mu$m, have angular resolutions of 8" (65\,kpc) and 13" (105\,kpc), respectively. The three SPIRE bands are 250, 350, and 500\,$\mu$m and have angular resolutions of 18" (150\,kpc), 25" (200\,kpc), and 36" (290\,kpc), respectively. The cutouts are provided in Fig. \ref{fig:cutouts}.

We further include an infrared spectrum from \textit{Spitzer}'s Infrared Spectrograph (IRS; \citealt{houck04}), which was presented in Fig. 1 of \cite{farrah10}. The IRS data have a spectral resolution of R $\sim57-126$ and an angular resolution of 5" (\mbox{40\,kpc}). Thus, unlike the SDSS data which are dominated by the emission from the quasar, the IRS and H-ATLAS data encompass the host galaxy as well and not just the nuclear region; as SDSS1214 is a type 1 quasar, its optical emission will be dominated by the nuclear emission, while the infrared data will include contributions from both the AGN and its host. The \textit{Herschel} beam could include neighbouring sources; however, given how bright the object is, the effect of this is likely to be small \citep{bethermin12, harris16}. Given SDSS1214's flux, any flux boosting is, at most, around ten per cent. 

We fit the SED of SDSS1214 using the pre-computed libraries of radiative transfer models for a starburst \citep{efstathiou00, efstathiou09}, as well as an AGN torus and polar dust \citep{efstathiour95, efstathiou13}, as in e.g. \cite{verma02, farrah07a, farrah12, farrah17, lonsdale15} \footnote{The models are publicly available at \href{http://ahpc.euc.ac.cy/index.php/resources/cygnus}{http://ahpc.euc.ac.cy/index.php/resources/cygnus}.}. The starburst models incorporate the stellar population synthesis model of \cite{bruzual93, bruzual03} and assume a Salpeter IMF. The model for the starburst has four parameters: the initial optical depth of the molecular clouds ($\tau_\textnormal{V}$), the starburst age (t$_*$), the e-folding time of the starburst ($\tau_*$), and the normalization factor which determines the starburst luminosity ($f_\textnormal{SB}$). We allow $\tau_\textnormal{V}$ to vary within the range 50--250, while we allow both the starburst age and e-folding time to vary within the range \mbox{10--30\,Myr}.

The AGN model is assumed to have a tapered geometry in which the thickness of the disc increases linearly with the distance from the central source out to some constant value, the reasons for which are discussed in \cite{efstathiou13}. It has five parameters: the equatorial optical depth of the torus ($\tau_\textnormal{UV}$), the ratio of the outer to inner torus radius ($r_2/r_1$), the half opening angle of the torus ($\Theta_0 = 90 - \Theta_1$), the inclination of the torus, and the normalization factor which determines the torus luminosity ($f_\textnormal{AGN}$). 

As we have no \textit{a priori} constraints on the parameters for either the starburst or AGN models, we use the full set of models and the Markov Chain Monte Carlo (MCMC) code SATMC \citep{johnson13}, which gives realistic uncertainties of the fitted parameters and derived physical quantities. 

We include the IRS data in the fit but reduce their spectral resolution so that they are matched to the spectral resolution of the radiative transfer models. The IRS data included in the fit with SATMC have a wavelength grid which is separated in steps of 0.05 in the log of the rest wavelength. Additional points around the 9.7$\mu m$ silicate feature and the PAH features are added to the equally spaced wavelength grid. The various fluxes and associated errors used in the SED fit are given in Table \ref{tab:flux}.

\begin{table}
	\begin{center}
    	\begin{tabular}{ccc}
        	\hline
            Source & $\lambda_\textnormal{obs}$ ($\mu$m) &  flux (mJy)\\
            \hline
             UKIDSS & 2.2 & $0.26 \pm 0.03 $   \\ 
             WISE & 3.4  &  $0.68 \pm 0.07$  \\
       		 & 4.6  &  $1.35 \pm  0.14$ \\
    	     & 12   & $3.05 \pm 0.31$  \\
  		     & 22   & $4.77 \pm 0.48$  \\         
             IRS& 8.14 &  $2.37 \pm  0.32$  \\     
  		     & 9.14 &  $2.73 \pm 0.35$  \\     
      		 & 10.25 &  $2.62 \pm 0.35$  \\    
      		 & 11.50 &  $3.11 \pm 0.70$  \\ 
             	 & \textit{11.66} &  $\textit{3.13} \pm  \textit{0.54}$  \\  
                 & \textit{12.68} &  $\textit{4.38} \pm  \textit{0.65}$  \\
     		 & 12.91 &  $3.74 \pm  0.72$  \\     
     		 & 14.48 &  $3.84  \pm 0.53$  \\ 
                  & \textit{15.75} &  $\textit{6.17} \pm  \textit{0.45}$  \\
     		 & 16.25 &  $5.79 \pm  0.50$  \\
             	  & \textit{17.59} &  $\textit{4.83} \pm  \textit{0.76}$  \\ 
     		 & 18.23 &  $2.69 \pm  0.79$  \\ 
                  & \textit{19.84} &  $\textit{3.80} \pm  \textit{0.96}$  \\ 
     		 & 20.46 &  $3.60 \pm 0.71$  \\     
     		 & 21.67 &  $3.87 \pm  0.63$  \\     
    	     & 22.95 &  $6.83 \pm  0.64$  \\ 
             	  & \textit{23.12} &  $\textit{6.89} \pm  \textit{0.59}$   \\ 
    	     & 24.31 &  $4.86 \pm  0.86$  \\     
      		 & 25.75 &  $5.87 \pm  0.53$  \\    
    	     & 27.28 &  $5.58 \pm  0.80$  \\     
     		 & 28.89 &  $5.74 \pm  0.67$  \\     
             MIPS  & 24   & $4.90 \pm  0.40$  \\ 
			 PACS & 100 & $214.45 \pm 41.32$ \\
			 & 160 & $163.96 \pm  47.95$ \\
			 SPIRE & 250 & $116.84 \pm  7.31$ \\
			 & 350 & $65.94 \pm  8.10$ \\
			 & 500 & $30.40 \pm  8.55$ \\
             \hline
    	\end{tabular}  
        \caption{Fluxes and associated errors used in the SED fit. The italicized IRS points are added to cover the positions of the PAH features and the 9.7$\mu$m silicate feature.}
        \label{tab:flux}
    \end{center}
\end{table}

\section{Results}\label{sec:results}

\subsection{CO line profile}\label{sec:co_profile}

\begin{figure}
    \begin{center}
        \includegraphics[width=\linewidth]{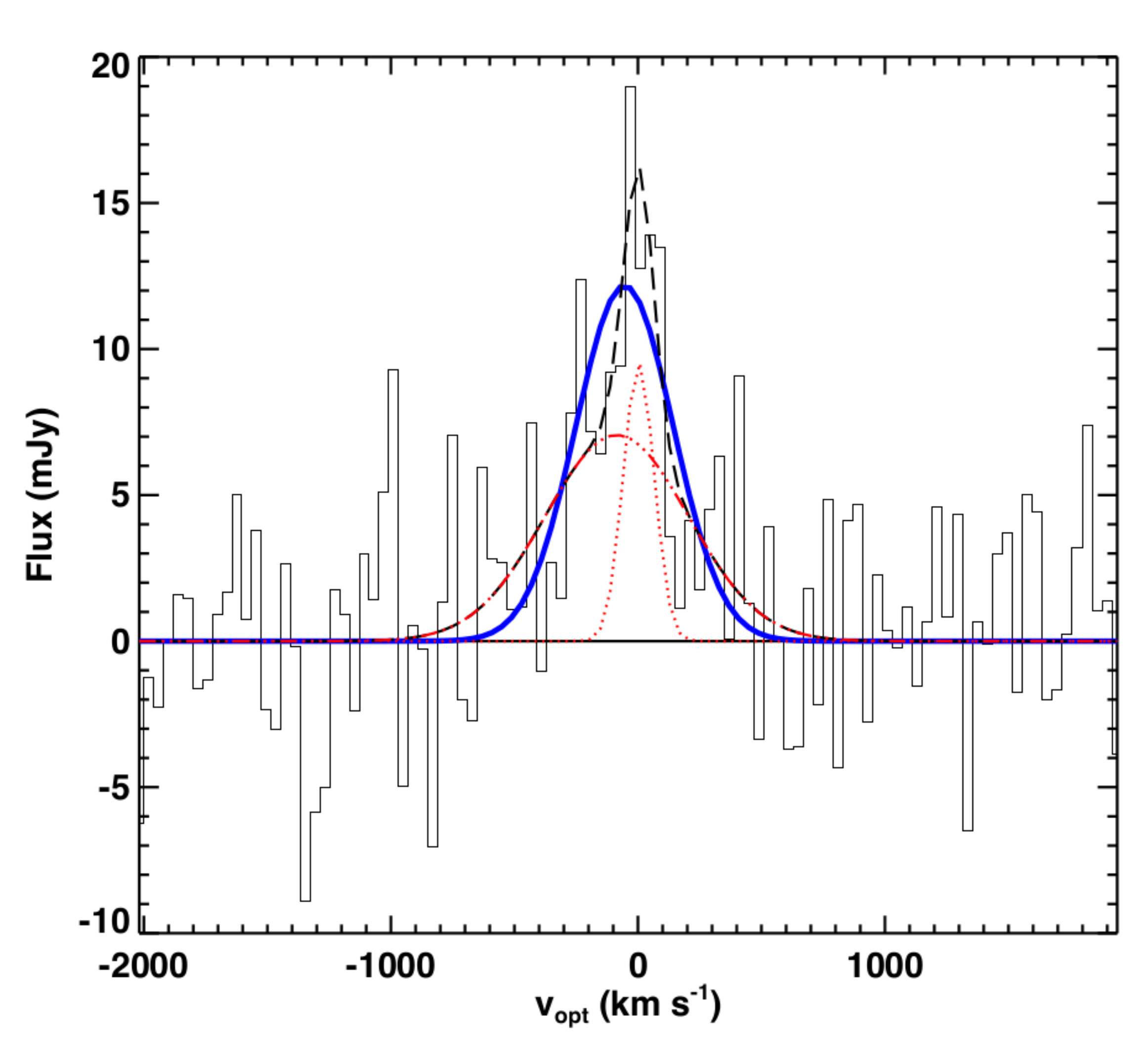}
        \caption{The CO \textit{J}=2--1 spectrum for SDSS1214, fit with both a single and double Gaussian. The velocities shown are all relative to systemic; thus, the bulk of the CO gas is moving at roughly systemic velocity. The single Gaussian is shown by the solid blue line. The total fit, the narrow component, and the broad component for the double Gaussian are shown by the dashed black, dotted red, and dash-dotted red lines, respectively. The parameter values resultant from the fits are given in the text. }
        \label{fig:gauss}
    \end{center}
\end{figure}

The one-dimensional CARMA spectrum (from $-2020 - +1940$\,km\,s$^{-1}$ in \mbox{40\,km\,s$^{-1}$} increments) is presented in Fig. \ref{fig:gauss}. It shows that the bulk of the CO gas is moving within \mbox{200\,km\,s$^{-1}$} of the systemic velocity. Furthermore, it appears to be consistent with a single narrow Gaussian profile. 

To quantify the CO emission, we fit a single Gaussian using \textsc{mpfit} \citep{markwardt09}, which performs a non-linear least squares fit. The single Gaussian fit is shown by the solid blue line in Fig. \ref{fig:gauss} and is characterized by an amplitude of \mbox{$12.2 \pm 1.2$\,mJy} centered about \mbox{$-40 \pm 20$\,km\,s$^{-1}$} with a width of \mbox{$200 \pm 20\,$\,km\,s$^{-1}$}. The fit is decent, with a reduced chi-squared value of 1.52.

For comparison, we also tried a double Gaussian (also shown in Fig. \ref{fig:gauss}) and a double-horned profile. The former gives an amplitude of \mbox{$9.5 \pm 2.7$\,mJy}, a center of \mbox{$0 \pm 20$\,km\,s$^{-1}$}, and a width of \mbox{$60 \pm 20$\,km\,s$^{-1}$} for the narrow component, while the broad component gives an amplitude of \mbox{$7.1 \pm 1.9$\,mJy}, a center of \mbox{$-90 \pm 60\,$km\,s$^{-1}$}, and a width of \mbox{$290 \pm 60$\,km\,s$^{-1}$}. The double-horned profile has a velocity width of \mbox{80\,km\,s$^{-1}$}.

The double Gaussian and double-horned fits return reduced chi-squared values of 1.46 and 1.50, respectively. Further, an F-variance test between the single and double Gaussian fits returns a significance value of 0.85, while that between the single Gaussian and double-horned profile returns 0.64. As these indicate that the fits do not have significantly different variances, neither the double Gaussian nor double-horned profiles are statistically preferred to the single Gaussian profile. Thus, we adopt the single Gaussian in our analysis.

\subsection{Infrared SED}\label{sec:ir_sed}
\begin{figure}
	\begin{center}
		\includegraphics[width=9cm]{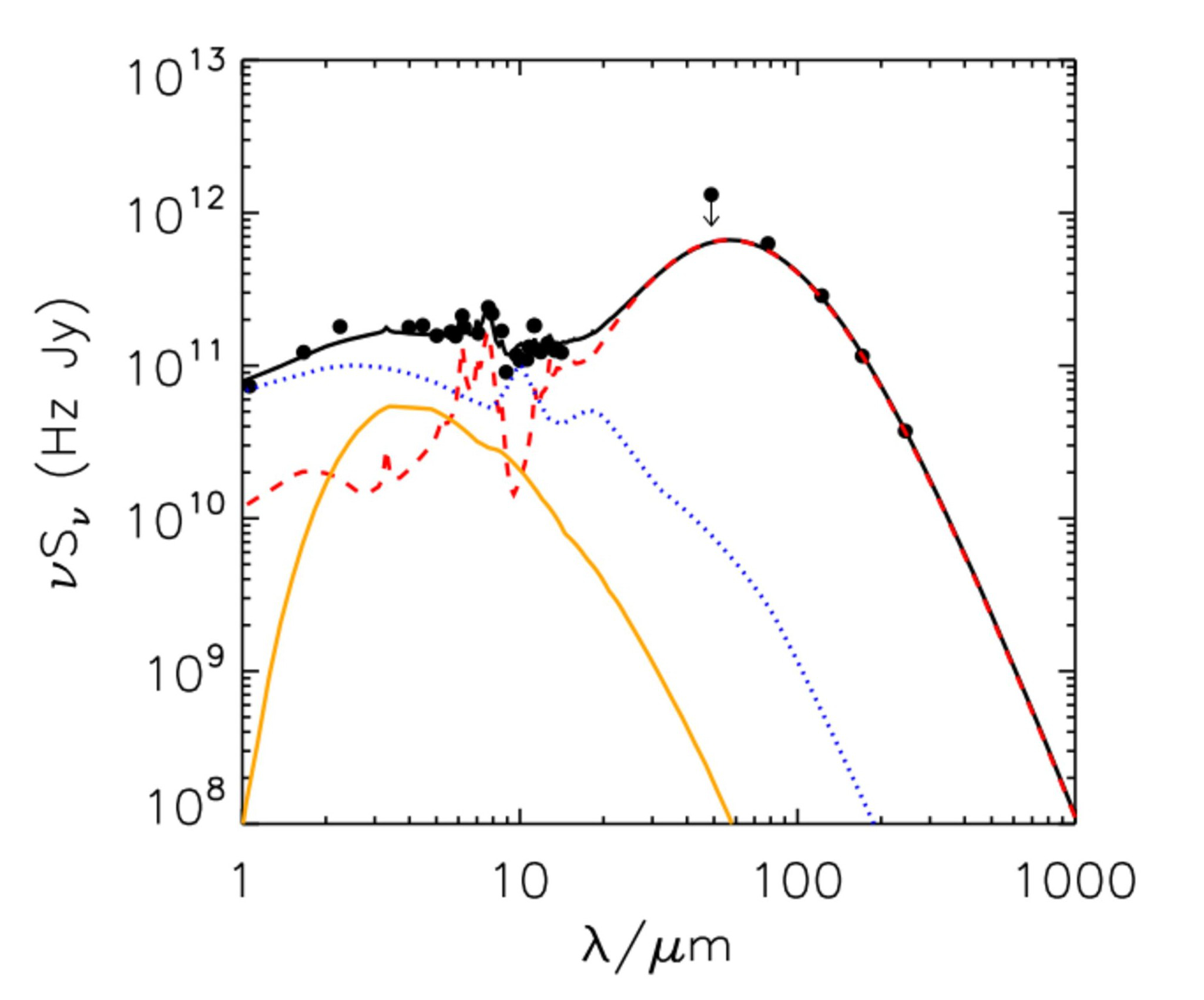}
		\caption{Rest-frame SED fit for SDSS1214. The total fit is depicted by the solid, black curve. The AGN and starburst components of the fit are shown by the blue dotted and red dashed lines, respectively. The orange curve shows the contribution from polar dust. The polar dust component allows for a better fit to the 3-5\,$\mu$m rest-frame spectrum.}
        \label{fig:sed}
    \end{center}
\end{figure}

The SED fit for SDSS1214 is shown in Fig. \ref{fig:sed}. The total rest-frame \mbox{1-1000\,$\mu$m} luminosity is \mbox{log($L_\textnormal{TOT}[\textnormal{L}_\odot]) = 13.04 \pm 0.04$} and is a combination of emission from the quasar, the starburst, and polar dust. The main results of the three component SED fit are provided in Table \ref{tab:ir_sed_prop}.

\begin{figure*}
	\begin{center}
	    \vspace*{-0.25cm}
	    \hspace*{-0.25cm}
		\includegraphics[width=1.02\linewidth, height=14cm]{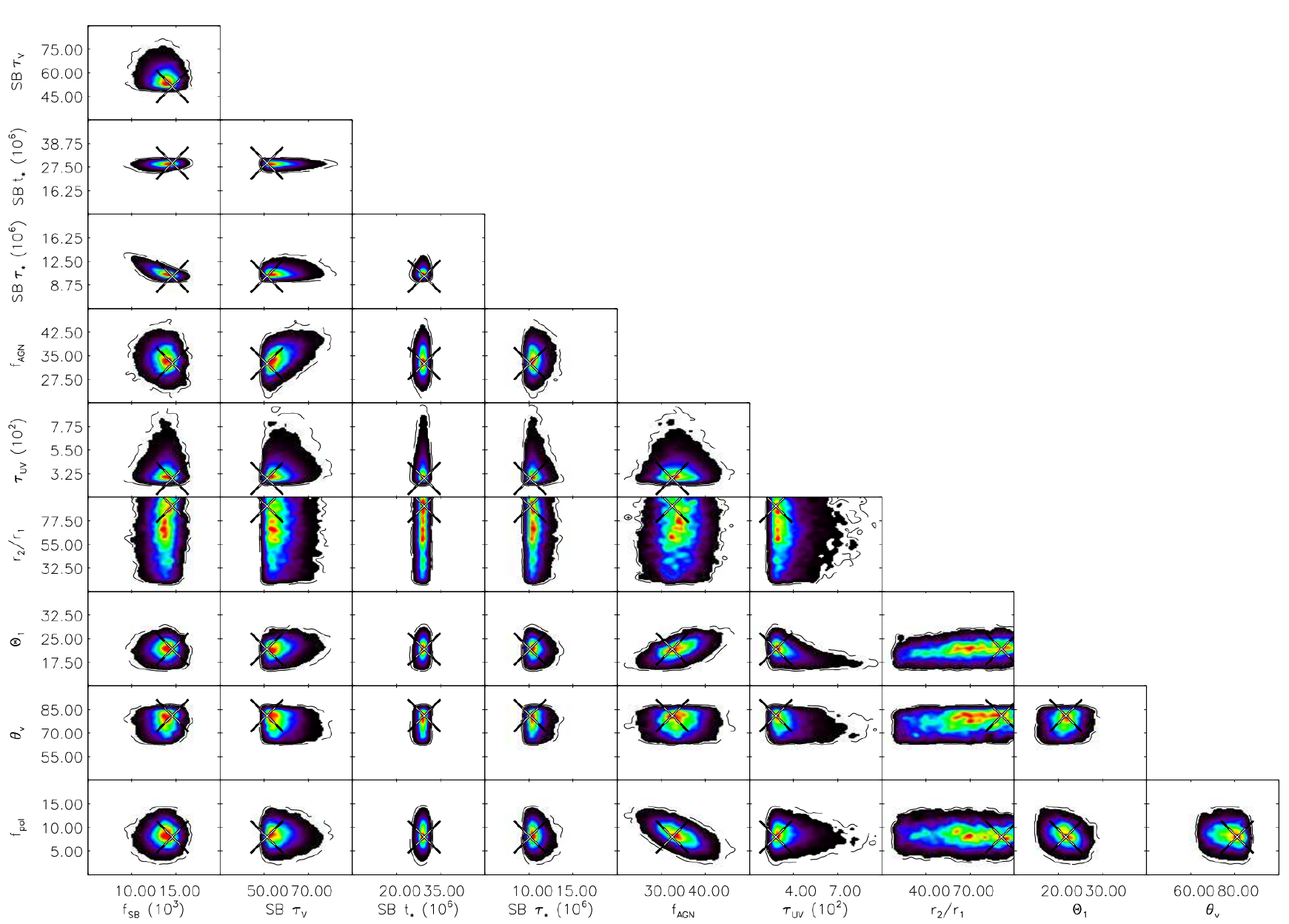}
		\caption{Covariances between the model parameters used in SDSS1214's SED fit described in Sec. \ref{sec:sed_analysis}. From this corner plot, it is clear that the four starburst parameters are well constrained by the fit. Within the set of AGN parameters, the ratio of the outer to inner torus radius is found to be degenerate. However, our conclusions are not impacted by this degeneracy.}
        \label{fig:corner}
    \end{center}
\end{figure*}

The emission is dominated by the starburst with a luminosity of \mbox{log($L_\textnormal{SB}[\textnormal{L}_\odot]) = 12.91 \pm 0.02$} and an initial optical depth of its molecular clouds of \mbox{$\tau_\textnormal{V} = 51.22^{+29.03}_{-0.22}$}. The SFR from the model fit (averaged over the last 50 Myr) is \mbox{2000$^{+40}_{-500}$\,M$_{\odot}$yr$^{-1}$}, and the age of the starburst is 30\,Myr \footnote{If we instead compute the SFR averaged over the age of the starburst, we find an SFR of \mbox{3300\,M$_{\odot}$yr$^{-1}$}.}. It is clear from the corner plot shown in Fig. \ref{fig:corner} that the four parameters of the starburst model (mentioned in Sec. \ref{sec:sed_analysis}) are well constrained by the fit. 

As PAH features are assumed to signify ongoing star formation \citep{peeters04}, they offer alternative estimates of the SFR. We use the 6.2$\mu$m feature, as it is one of the most prominent PAH features in SDSS1214's spectrum and does not require a deblending against the [S\begin{scriptsize}{IV}\end{scriptsize}] line at 10.5$\mu$m, as the 11.3$\mu$m feature would. To estimate the 6.2$\mu$m flux, we rely on the IRS spectral decomposition code \textsc{pahfit} \citep{smith07}.

There exist several conversions between PAH luminosities and SFRs \citep[e.g.][]{peeters04, wu05, pope08, lutz08, hernan09, shipley16}. We adopt the \cite{shipley16} relation even though their calibration sample was located at lower redshifts than that of SDSS1214 because their relations were shown to accurately reproduce a wider variety of observations\footnote{Many of the studies that use PAH luminosities to infer SFRs use a scaling of the well-known FIR-derived SFR of \cite{kennicutt98}. For SDSS1214, we find an SFR of \mbox{$1400$\,M$_{\odot}$yr$^{-1}$}, which is lower than our SED- and PAH-derived SFRs. However, as errors on the FIR-derived value are difficult to estimate, it is still possible that all of these values align within the errors. For a discussion on the SFR estimators in quasars, see \cite{feltre13}}. Thus, from the luminosity of this feature, we can infer an SFR using the following:

\begin{equation}
	\textnormal{log(SFR)} = (-41.73 \pm 0.08) + \textnormal{log}(L_{6.2\mu\textnormal{m}}),
\end{equation}

\noindent where the SFR is given in units of M$_{\odot}$yr$^{-1}$, and the luminosity is in erg\,s$^{-1}$. We find an SFR of \mbox{$2000 \pm 600$\,M$_{\odot}$yr$^{-1}$} and thus, the PAH-derived SFR aligns nicely with that from the SED fit\footnote{Alternatively, using the 11.3$\mu$m PAH feature and its respective SFR relation gives an SFR of \mbox{$2400 \pm 1000$\,M$_{\odot}$yr$^{-1}$}, which is still consistent with our SED-derived rate within the errors.}.

The AGN has an infrared luminosity of \mbox{log($L_\textnormal{AGN}[\textnormal{L}_\odot]) = 12.36^{+0.14}_{-0.15}$}, or 21 per cent of the total infrared emission; this luminosity is that of the full AGN component, i.e. the primary emission from the source and the reprocessed infrared emission. In degrees, \mbox{$\Theta_0 = 68.32^{+5.67}_{-7.08}$}, while the viewing angle measured from the equatorial plane (again in degrees) is \mbox{$\theta_\textnormal{v} = 81.07^{+3.92}_{-16.06}$}. The inclination of the torus is then given as \mbox{$90 - \theta_\textnormal{v}$}. That \mbox{$\theta_\textnormal{v}> 90 - \Theta_0$} is expected since SDSS1214 is a Type 1 quasar in which the broad-line region is directly visible. Additionally, \mbox{$\tau_\textnormal{UV} = 280^{+780}_{-20}$}. Furthermore, the visual extinction, $A_V$, is $\sim\tau_\textnormal{UV}/5$, while the optical depth at 9.7$\mu$m is $\tau_{9.7\mu\textnormal{m}}\sim \tau_\textnormal{UV}/61$ \citep{rowan92}.
The ratio of the outer to the inner radius is \mbox{$r_2/r_1 = 91.42^{+7.52}_{-70.41}$}; $r_2/r_1$ is degenerate, as this parameter mainly determines the long wavelength tail of the torus emission, which is poorly constrained in the fitting since the starburst dominates the emission within this wavelength range. The conclusions of this paper are not affected by this degeneracy.

In addition to the equatorial torus, several studies have shown that a component along the polar axis is required to describe the observed MIR emission in AGN \citep[e.g.][]{honig12,honig13, tristram14, lopez16}, with this polar component often being brighter in the MIR than the torus \citep[for a schematic of such a system, see Fig. 1 of][]{ramos17}. It is proposed that this polar dust originates from a dusty wind within the inner region of the torus that is blown out into the narrow-line region \citep{honig12, honig13, tristram14}. This is a similar model to that proposed by \cite{efstathiou95} for NGC1068 and \cite{efstathiou06} for the hyperluminous infrared galaxy IRASF 10214+4724. More recently, a model for polar dust has been used for the interpretation of the dust-enshrouded tidal disruption event in Arp 299 \citep{mattila18}. Five per cent of the total infrared emission comes from this polar dust, which has a luminosity of \mbox{log($L_\textnormal{dust}[\textnormal{L}_\odot]) = 11.75^{+0.26}_{-0.46}$} and consists of optically thick clouds at a constant temperature of 900\,K. While the polar dust accounts for a lower percentage of the total infrared emission than does the torus (5 versus 21 per cent), it does make a contribution of $\sim$30 per cent to the 3-5\,$\mu$m rest-frame spectrum, where it is required to fit the SED at much higher significance. The fit in the near-infrared is thus much improved over the SED presented in \cite{farrah12}. 

We have not explored varying the polar dust temperature with the MCMC code, but have allowed the normalization factor which controls its luminosity to vary. We instead have fit the infrared data several times assuming different discrete values of the polar dust temperature in the range \mbox{700 -- 1100\,K}. We find that a value of 900\,K provides the best fit and that the rest of the model parameters are not significantly affected by the choice of the polar dust temperature.

While the addition of the polar dust component does allow for a better fit to the data in the 3-5$\mu$m range, it does not have a significant effect on the output luminosities; the results from a two component fit (AGN and starburst) give a total rest-frame \mbox{1-1000\,$\mu$m} luminosity of \mbox{log($L_\textnormal{TOT}[\textnormal{L}_\odot]) = 13.04 \pm 0.04$}, an AGN luminosity of \mbox{log($L_\textnormal{AGN}[\textnormal{L}_\odot]) = 12.43 \pm 0.09$}, and a starburst luminosity of \mbox{log($L_\textnormal{SB}[\textnormal{L}_\odot]) = 12.93^{+0.02}_{-0.04}$}. This gives an SFR of \mbox{2000$^{+90}_{-500}$\,M$_{\odot}$yr$^{-1}$}. As these values are comparable to those from the three-component fit, the inclusion or exclusion of the polar dust component in the SED fit does not affect the main results of this paper.

\begin{table}
    \renewcommand*{\arraystretch}{1.4}
    \centering
    \begin{tabular}{cc}
        \hline
        \hline
        \multicolumn{2}{c}{IR SED properties}  \\
        \hline
        log($L_\textnormal{TOT}$[L$_\odot$]) & $13.04 \pm 0.04$ \\
        log($L_\textnormal{AGN}$[L$_\odot$]) & $12.36^{+0.14}_{-0.15}$ \\
        log($L_\textnormal{SB}$[L$_\odot$]) & $12.91 \pm 0.02$ \\
        log($L_\textnormal{dust}$[L$_\odot$]) & $11.74^{+0.26}_{-0.47}$ \\
         SFR (M$_\odot$yr$^{-1}$) & $2000^{+40}_{-500}$  \\
        \hline
        \hline
    \end{tabular}
    \caption{IR SED properties. The SED was found to comprise of three components: an AGN torus, polar dust, and a starburst. }
    \label{tab:ir_sed_prop}
\end{table}

\section{Discussion}\label{sec:discussion}

\subsection{Molecular gas properties} \label{sec:co_prop}

For SDSS1214, the CO line luminosity, as defined in \cite{solomon05}, is \mbox{$9.2 \times 10^{10}$\,K\,km\,s$^{-1}$pc$^2$}. This can be converted into a gas mass using the relation (as in e.g. \citealt{solomon05} and \citealt{bolatto13}):

\begin{equation}
    M_\textnormal{gas} = \alpha L'_\textnormal{CO},
\end{equation}

\noindent where $M_\textnormal{gas}$ includes both the H$_2$ and He masses and is expressed in units of M$_\odot$, $\alpha$ is the CO-to-H$_2$ conversion factor, and $L'_\textnormal{CO}$ is the CO luminosity.

Within the Milky Way, \cite{bolatto13} estimate $\alpha$ to be \mbox{4.3 M$_\odot$ (K\,km\,s$^{-1}$pc$^2$)$^{-1}$}. However, \cite{downes98} find the factor to be much lower for ULIRGs, at \mbox{0.8 M$_\odot$ (K\,km\,s$^{-1}$pc$^2$)$^{-1}$}. Using the \cite{bolatto13} conversion factor gives a gas mass of \mbox{$M_\textnormal{gas} = 3.9 \times 10^{11}\,\textnormal{M}_\odot$}, while that from \cite{downes98} gives \mbox{$M_\textnormal{gas} = 7.3 \times 10^{10}\,\textnormal{M}_\odot$}. Given SDSS1214's starburst luminosity and subsequent ULIRG classification, we adopt the latter, lower value for the gas mass. A summary of the CO line properties and subsequent derived quantities is given in Table \ref{tab:co_properties}.

\begin{table}
    \renewcommand*{\arraystretch}{1.4}
    \centering
    \begin{tabular}{cc}
        \hline
        \hline
        \multicolumn{2}{c}{CO line profile}  \\
        \hline
        amplitude (mJy)   & $12.2 \pm 1.2$ \\
        center (km\,s$^{-1}$) & $-40 \pm 20$ \\
        width (km\,s$^{-1}$) & $200 \pm 20$ \\
        \hline
        \hline
        \multicolumn{2}{c}{Derived quantities} \\
        \hline
        $L'_\textnormal{CO}$ (K\,km\,s$^{-1}$pc$^2$) & $9.2 \times 10^{10}$ \\
        $M_\textnormal{gas}$ (M$_\odot$) & $7.3 \times 10^{10}$ \\
        $t_\textnormal{gas}$ (Myr) & 40 \\
        \hline
        \hline
    \end{tabular}
    \caption{CO line profile and derived quantities. The parameters of the CO line profile describe the single Gaussian. The method for computing the derived quantities is given in Sec. \ref{sec:co_prop}. The gas-depletion time-scale, $t_\textnormal{gas}$, is $M_\textnormal{gas}$/SFR, where SFR takes the value given in Table \ref{tab:ir_sed_prop}.}
    \label{tab:co_properties}
\end{table}

Because we assume no lensing, our gas mass estimate serves as an upper limit; following \cite{solomon05}, the mass takes the form of $7.3\mu^{-1} \times 10^{10} \textnormal{M}_\odot$ and thus any magnification would lower our gas mass estimate. Within our data, we see no evidence for lensing, but we would need HST imaging to completely rule out any lensing.

The molecular gas mass of SDSS1214 is comparable to the masses of higher redshift objects, such as those in the \cite{riechers11} sample, in which all objects reside at $z > 2$. Within this sample, the authors find \mbox{$M_\textnormal{gas} =(0.27 - 14.7) \times 10^{10}\,\textnormal{M}_\odot$}. Further, Fig. 9 of \cite{carilli13} show the far-infrared luminosity as a function of $L'_\textnormal{CO}$ for various types of objects, including quasars, submillimetre galaxies, and radio galaxies. SDSS1214 does align nicely (albeit at the upper-end of the relation) with the other quasars. As these are also generally at higher redshifts though, SDSS1214 might appear unusual given its redshift. However, because FeLoBALs are obscured, optical quasar searches are biased towards high luminosity systems and when instead comparing SDSS1214 to other objects with similar AGN luminosities (rather than redshifts), we again find similar molecular gas masses \citep[e.g.][]{iono06, brusa15, fiore17}.

We now look at the CO emission in individual velocity slices. The channel maps in Fig. \ref{fig:co_chanmaps} show the CO \mbox{\textit{J} = 2--1} emission in a subset of these velocities, \mbox{-470 -- +130\,km\,s$^{-1}$}. Nearly seventy per cent of the CO emission is located within \mbox{-229 -- +88\,km\,s$^{-1}$} and \mbox{30\,kpc} of the quasar. The mass of the gas within this velocity range is \mbox{$M_\textnormal{gas} = 4.8 \times 10^{10}\,\textnormal{M}_\odot$}. The confinement in location and spread in velocities are consistent with the CO being located within the host galaxy of the quasar. The small excess of gas at negative velocities could signpost a weak outflow along our line of sight, but is more plausibly interpreted as arising from some combination of the dynamics and relative orientation of the host galaxy. There is also a weak ($\lesssim$3.5$\sigma$) detection of CO gas \mbox{30\,kpc} from the quasar in both the -430 and \mbox{-390\,km\,s$^{-1}$} velocity slices. This could indicate a weak outflow, but as the signal-to-noise ratio is low, we do not remark on it further.

\color{Black}

\subsection{AGN \& starburst properties}

\begin{figure}
	\begin{center}
        \includegraphics[width=1.1\linewidth]{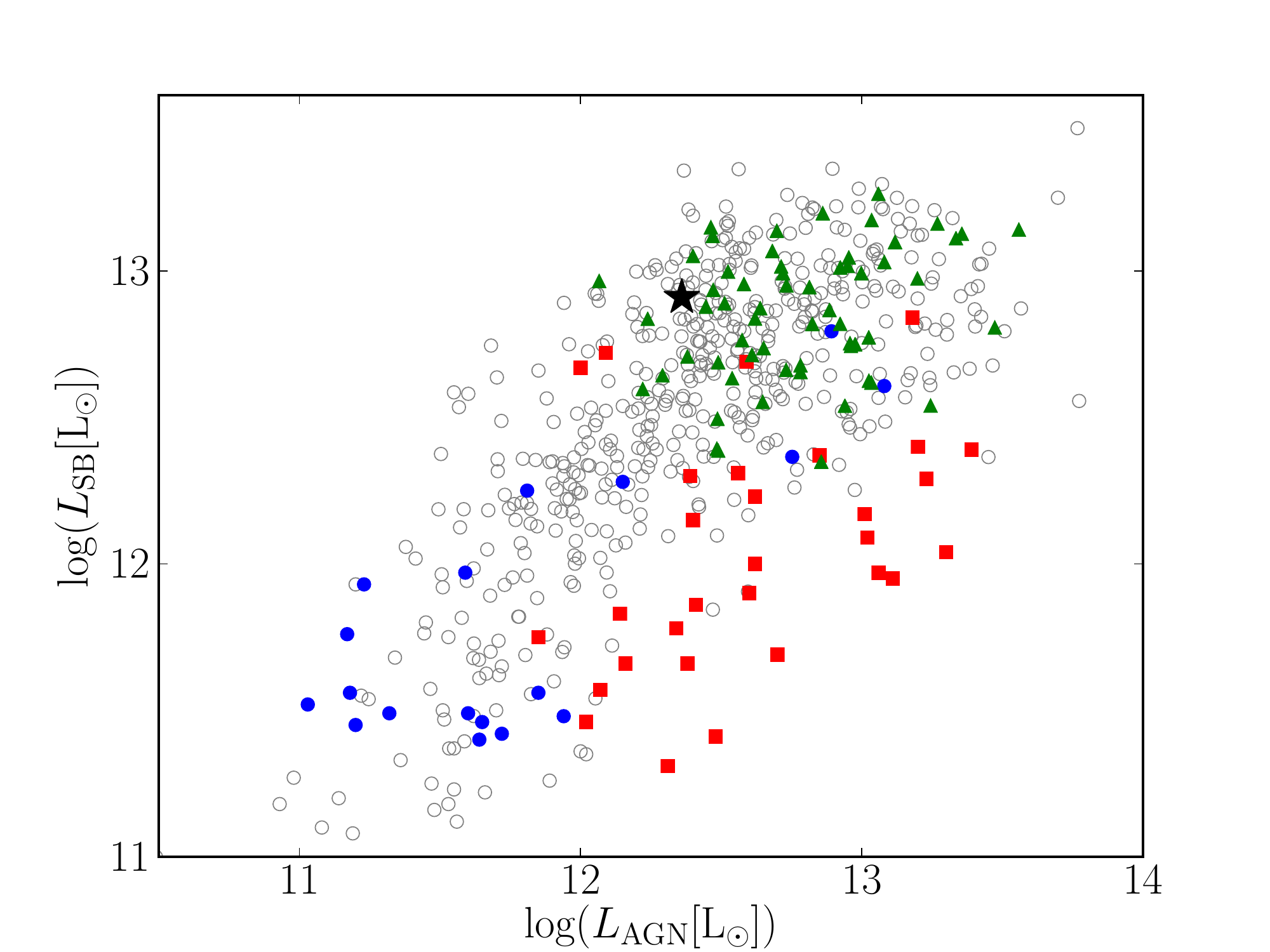}
        \caption{Comparison between SDSS1214 and other studies in terms of AGN and starburst luminosities. SDSS1214 is the black star. Other FeLoBALs are shown by the red squares. Blue circles and green triangles correspond to LoBALs and HiBALs, respectively. The HiBALs are individually detected in the FIR and thus exhibit much higher SFRs than would `typical' HiBAL quasars. Non-BAL quasars are the open circles. SDSS1214 shows a comparable $L_\textnormal{AGN}$ to the other (Fe)LoBALs, but displays a higher $L_\textnormal{SB}$.}
        \label{fig:farrah_comp}
    \end{center}
\end{figure}

\cite{farrah12}, hereafter F12, have previously compiled a sample of 31 FeLoBAL quasars, including SDSS1214. Our value for SDSS1214's starburst luminosity is higher than theirs, as we have the \textit{Herschel} photometry to help constrain its starburst properties. However, our estimate for the AGN luminosity aligns with theirs within the errors. When considering SDSS1214 along with the remaining F12 objects, we find SDSS1214 to have a comparable AGN luminosity, but to be much more star-forming. In fact, SDSS1214 is the most luminous star-forming FeLoBAL known. Fig. \ref{fig:farrah_comp} shows SDSS1214 (black star) in relation to the rest of the F12 sample (red squares). 

\cite{lazarova12} have studied a sample of lower redshift LoBALs (shown by the blue circles in Fig. \ref{fig:farrah_comp}) and have found that the SFRs of the IR-luminous LoBALs are eighty per cent higher than those of non-BALs. Additionally, the studies of \cite{cao12} and \cite{pitchford16}, hereafter CO12 and P16, respectively, have shown the HiBAL population to be indistinguishable from the non-BAL population. Thus, we would expect higher levels of star formation in our FeLoBAL than in HiBALs/non-BALs with similar properties (i.e. redshift and/or AGN luminosity). CO12 largely studied objects at higher redshifts than SDSS1214's that were individually undetected by SPIRE. The authors therefore relied on a stacking analysis to obtain SFR estimates. These objects were found to have SFRs of \mbox{$\sim250$\,M$_\odot$yr$^{-1}$}, much lower than the SFR of SDSS1214. 

P16, on the other hand, only studied objects that were individually detected by SPIRE and thus more star-forming than the general quasar population. The P16 HiBAL objects are represented by the green triangles in Fig. \ref{fig:farrah_comp}, while the P16 non-BAL quasars are the open circles. When considering P16's objects with similar AGN luminosities to SDSS1214's, the mean SFR is 1200\,M$_{\odot}$yr$^{-1}$ and 1000\,M$_{\odot}$yr$^{-1}$ for the HiBALs and non-BALs, respectively. While both values are lower than that found for SDSS1214, they do align nicely when allowing for the eighty per cent increase between the SFRs of HiBALs/non-BALs and LoBALs. Thus, in terms of star formation, SDSS1214 is not a typical quasar, but does align with other extremely star-forming quasar systems.

The SFR and gas mass give a starburst time-scale of 40\,Myr, assuming the lower conversion factor of \cite{downes98}. This starburst time-scale is lower than those found in the works of \cite{feruglio10} and \cite{alatalo15}, both of which studied Mrk 231. The latter gives a time-scale of 110\,Myr, assuming that star formation within Mrk 231's host is solely responsible for depleting the central gas, just as we are with SDSS1214's host. However, because our SFR is significantly higher, our time-scale is lower. \cite{yan10} found similar depletion time-scales in their sample of ULIRGs (17 -- 96\,Myr with an average value of 38\,Myr). Additionally, \cite{cicone14} studied a sample of low-$z$ ULIRGs and found that the time-scale for gas depletion due to star formation is $10^7$--$10^9$\,yr, with quasars at the lower end of the range and starburst galaxies at the upper limit.

Our 40\,Myr time-scale assumes not only that star formation is the sole mechanism responsible for depleting the gas in the system, but also that the star formation will continue at a constant rate of \mbox{2000\,M$_{\odot}$yr$^{-1}$}. However, because the starburst model used actually assumes an exponentially decaying SFR, this starburst time-scale serves instead as a lower limit. 

Finally, we briefly remark on the issue of AGN feedback. Signatures of molecular outflows, which would serve as evidence for ongoing AGN feedback, are typically found via broad, faint wings and/or asymmetries in the CO line \citep[e.g.][]{feruglio10, alatalo11, cicone14, veilleux17}. While we do not see such signatures in our data, the CARMA spectrum is of too low signal-to-noise for them to be visible. We can thus draw no conclusions on the presence or absence of AGN feedback in SDSS1214. It is, however, notable that we see no evidence for a double-peaked CO line profile; thus, we do not find evidence in the CARMA data to support a merger \citep[see also e.g.][]{yan10}. Moreoever, the CO line properties of SDSS1214 - a narrow Gaussian centered close to the systemic velocity - are consistent with the CO line properties of the classical quasars presented by \cite{riechers11}. Therefore, while we cannot rule out AGN feedback, our results can plausibly be interpreted as the CO gas being at least mostly associated with the star-forming regions, rather than the BAL wind. However, even if we assume no ongoing feedback in SDSS1214, this does not rule out AGN feedback occurring at some point in this system, since the starburst time-scale is likely much longer than potential feedback time-scales \citep[e.g.][]{cicone14}.

\section{Conclusions} \label{sec:conclusions}

We have presented CARMA CO observations of SDSS1214, an FeLoBAL at $z=1.046$. Our data specifically describe the \mbox{$J$=2--1} transition and thus the cold CO gas. We have additionally compiled the optical data (taken from SDSS), as well as all available infrared data in order to obtain the best possible fit to the SED. 

The SED fit attributes the total luminosity of the system to a combination of the AGN torus, a starburst, and polar dust. The system is dominated by emission from the starburst, which is better constrained than in previous studies thanks to FIR data from H-ATLAS; 74 per cent of the total infrared emission (from 1-1000$\mu$m) is attributable to the starburst component. The torus contributes 21 per cent of the total emission, while the polar dust contributes five per cent. The inclusion of this third component does not largely affect the various output parameters, but does allow for a much better fit to the 3-5$\mu$m rest-frame spectrum.

The CO profile is found to be reproduced equally well by a single Gaussian, a double Gaussian, and a double-horned profile. Due to the lack of statistical difference among the various options, we assume the single Gaussian fit. The CO data imply a gas mass of $7.3 \times 10^{10}$\,M$_\odot$. The SFR of \mbox{2000$^{+40}_{-500}$\,M$_{\odot}$yr$^{-1}$} is taken from the SED fit and is further confirmed using the 6.2$\mu$m PAH-derived SFR. The gas mass and SFR then give a gas depletion time-scale (due only to star formation) of 40\,Myr. 

Given this FeLoBAL's redshift, it exhibits an extremely high SFR; in fact, it is harbours the most luminous starburst of any FeLoBAL quasar known. While this still holds true when considering its AGN luminosity, it does better align with the rest of the extreme star formation quasar population. SDSS1214 is shown to have similar IR properties to other classes of quasars and as such, its SFR seems to fall on the upper end of the range of SFRs observed in FIR-selected quasars. 

Lastly, we briefly examine evidence for the existence of AGN feedback within this system. If the AGN was quenching the star formation, then we may expect to see a significant fraction of the CO gas at a similar redshift to the BAL wind, rather than to the PAHs. Conversely, if the BAL wind was triggering star formation, we might expect to see some PAH emission at a similar velocity to the BAL wind. Because the CO, PAHs, and narrow emission lines are consistent with the same systemic velocity apart from the BAL wind, neither the star formation nor the CO gas that fuels it appear to be associated with the BAL wind. Thus, it seems most plausible that the CO gas and the star formation are currently unaffected by the presence of the quasar wind. This could be due to the scale of the BAL wind itself; it is possible that the BAL wind only extends out from the central source on sub-kpc scales and therefore cannot reach the CO gas.

\section*{ACKNOWLEDGEMENTS}

Support for CARMA construction was derived from the Gordon and Betty Moore Foundation, the Kenneth T. and Eileen L. Norris Foundation, the James S. McDonnell Foundation, the Associates of the California Institute of Technology, the University of Chicago, the states of California, Illinois, and Maryland, and the National Science Foundation. Ongoing CARMA development and operations are supported by the National Science Foundation under a cooperative agreement, and by the CARMA partner universities. JA acknowledges financial support from the Science and Technology Foundation (FCT, Portugal) through research grants PTDC/FIS-AST/2194/2012, PTDC/FIS-AST/29245/2017, UID/FIS/04434/2013, and UID/FIS/04434/2019.

\label{lastpage}

\end{document}